\documentclass[12pt]{iopart}
 \usepackage{graphicx}
\begin{document}

\title[High-field magnetization of LiV$_2$O$_4$]{High-field magnetization
of the 3$d$ heavy-fermion system LiV$_2$O$_{4-\delta}$ ($\delta$ = 0, 0.08)
}

\author{N Tsujii\dag, K Yoshimura\ddag, K Kosuge\ddag,
H Mitamura\S\ and T Goto\S, 
}

\address{\dag\ National Institute for Materials Science,
Sengen 1-2-1, Tsukuba, 305-0047, Japan}

\address{\ddag\ Graduate School of Science, Kyoto University, Sakyo-ku,
Kyoto, 606-8502, Japan}

\address{\S\ Institute for Solid State Physics, University of Tokyo,
5-1-5 Kashiwanoha, Kashiwa, 277-8581, Japan}

\begin{abstract}
Metamagnetic behavior has been observed in
LiV$_2$O$_4$ powder sample around 38 T at 4.2 K.
On the other hand, magnetization for oxygen deficient LiV$_2$O$_{3.92}$ 
shows no indication of metamagnetism up to 40 T, and shows
substantially reduced magnetic moment compared to that of LiV$_2$O$_4$. 
These results suggest that ferromagnetic interaction 
is strongly enhanced by magnetic fields in LiV$_2$O$_4$, whereas
antiferromagnetic interaction is dominant in LiV$_2$O$_{3.92}$.
\end{abstract}

\pacs{71.27.+a, 75.40.Cx, 75.90.+w}

\submitto{\JPCM}

\ead{TSUJII.Naohito@nims.go.jp}
%\maketitle

\section{Introduction}
Recently, research on the spinel compound LiV$_2$O$_4$ 
has been an exciting field since the discovery
of the large electronic specific heat coefficient
$\gamma$ = 420 mJ/mol$\cdot$K$^2$ at low temperature ~\cite{Kondo}.
This value of $\gamma$ is the largest among $d$-electron compounds.
Moreover, LiV$_2$O$_4$ has been found to exhibit heavy-fermion like
behavior; 
i.e., a broad maximum in the magnetic susceptibility $\chi$~\cite{Kondo,Onoda,Ueda},
$T^2$-dependence in the electrical resistivity~\cite{Urano},
and the Korringa type relaxation, $1/T_{1} \propto T$, where $1/T_1$ is
the spin-lattice relaxation rate derived from the $^7$Li-NMR measurements
~\cite{Kondo,Fujiwara,Mahajan}.
It should also be noted that the solid-solution Li$_{1- x}$Zn$_x$V$_2$O$_4$
undergoes a metal-insulator transition as a function of 
$x$~\cite{Onoda,Kawakami,Muhtar}.
In spite of enormous studies, however, origin of these anomalous
behavior is not clearly solved.

Several models have been proposed to explain the 
physical properties of LiV$_2$O$_4$.
At first, the large $\gamma$ has been attributed to the Kondo-lattice
formation as in the case of $f$-electron based compounds
~\cite{Kondo}.
Mahajan \textit{et al.} have shown that their $^7$Li-NMR results 
are consistent with this mechanism.
However, temperature dependence of the electrical resistivity
at high temperatures differs qualitatively from that of the $f$-electron based
Kondo-lattice systems~\cite{Urano}.
For the Kondo-lattice formation in LiV$_2$O$_4$, both the conduction
and the localized electrons are needed, whereas 
the presence of well-defined localized electrons is not clearly demonstrated.
On the other hand, Fujiwara \textit{et al.}
have interpreted their $^7$Li-NMR results 
as the characteristic behavior of itinerant-electron systems
with strongly-enhanced ferromagnetic spin-fluctuations~\cite{Fujiwara}. 
They have shown 
that the temperature dependence of $1/T_1$ 
is well described by the spin-fluctuation theory (SCR theory)
for weakly- or nearly-ferromagnetic metals
~\cite{Fujiwara}.
Neutron scattering studies by Krimmel {\it et al.} have also suggested
possible nearly-ferromagnetic state for LiV$_2$O$_4$ at high temperatures
~\cite{Krimmel}.
Thus, physical properties of LiV$_2$O$_4$ have been argued 
in terms of completely
different electronic states, i.e., localized- or itinerant-electron systems.
In addition, the effect of geometrical frustration 
arising from the spinel structure is suggested to be essential for
the anomalous behavior in LiV$_2$O$_4$~\cite{Urano,Lacroix}.
This is inferred from the observations of spin-glass order in a slightly 
Mg- or Zn-doped variants~\cite{Onoda,Ueda,Urano}.

To obtain further insight for the magnetic state of LiV$_2$O$_4$, 
we have performed a high-field magnetization measurement on LiV$_2$O$_4$.
Since magnetic field can enhance the ferromagnetic spin-correlation,
this experiment would give valuable information about the spin
correlation in LiV$_2$O$_4$.
To see the effect of geometrical frustration, we have also studied on
an oxygen-deficient sample LiV$_2$O$_{3.92}$.

\section{Experimental}

Powder sample of LiV$_2$O$_4$ was prepared by a solid state reaction 
from powder samples of Li$_3$VO$_4$, V$_2$O$_3$, and V$_2$O$_5$.
The same procedure reported in ref.~\cite{Nakajima} was employed.
An oxygen deficient sample LiV$_2$O$_{3.92}$ was prepared by varying the 
nominal composition. X-ray diffraction confirmed the spinel structure 
for both the samples. Magnetic susceptibility was measured 
using a superconducting quantum interference device
(SQUID) magnetometer at 1 kOe. High-field magnetization measurements were performed 
at 4.2 K by an induction method with well-balanced pickup coils. 
Magnetic fields up to 45 T were generated by a long-pulse magnet 
at the Institute for Solid State Physics, University of Tokyo.

\section{Results}
Figure 1 shows inverse magnetic susceptibility 1/$\chi$ of LiV$_2$O$_4$ 
and LiV$_2$O$_{3.92}$ as functions of temperature $T$.
%%%%%%%%%%%%%%%%%%%%%%%%%%%%%%%%%%%%%%%%%%%%%%%%%%%%%%%%%%
%\begin{figure}[!tbp]
% \begin{center}
% \includegraphics[width=10cm]{Fig1.eps}
% \end{center}
% \caption{
% Inverse of the magnetic susceptibility $1/\chi$ of LiV$_2$O$_4$ and LiV$_2$O$_{3.92}$
% powder samples. Dotted line represents the corrected data of LiV$_2$O$_4$ 
% with the Curie term
% subtracted.}
%\end{figure}%
%%%%%%%%%%%%%%%%%%%%%%%%%%%%%%%%%%%%%%%%%%%%%%%%%%%%%%%%%%%%
For both the systems, $\chi$ follows a Curie-Weiss behavior 
at high temperatures, and shows no evidence of magnetic ordering
down to 2 K. 1/$\chi$ of LiV$_2$O$_4$ shows 
a saturation behavior around 10 K, and slightly decreases below 5 K. 
This small decrease is attributed to the contribution of 
impurities and/or defects. Subtracting this contribution
by a Curie term yields a minimum at $T_{\rm m}$ = 16 K,
as can be seen in the dotted line in Fig. 1.
This is consistent with those reported by several authors
~\cite{Kondo,Onoda,Ueda}.

1/$\chi$ of LiV$_2$O$_{3.92}$ also shows a Curie-Weiss behavior, 
and the data above 60 K agree with those of LiV$_2$O$_4$.
Below about 20 K, 1/$\chi$ of LiV$_2$O$_{3.92}$ 
decreases rapidly without any anomaly.
Absence of anomaly in $\chi$ of LiV$_2$O$_{3.92}$ has been
suggested previously~\cite{Onoda}.
One may consider that a large contribution from defects
masks a minimum in the intrinsic $1/\chi$ of LiV$_2$O$_{3.92}$.
However, we found that a minimum is not given in $1/\chi$ of LiV$_2$O$_{3.92}$ 
by reasonable corrections.
This suggests that $\chi$ of LiV$_2$O$_{3.92}$ at low temperature is qualitatively
different from that of LiV$_2$O$_4$, though the high-temperature
behavior agrees each other.

Figure 2 shows magnetization $M$ of LiV$_2$O$_4$ 
and LiV$_2$O$_{3.92}$ as functions of magnetic field $H$.
%%%%%%%%%%%%%%%%%%%%%%%%%%%%%%%%%%%%%%%%%%%%%%%%%%%%%%%%%%%
%\begin{figure}[!tbp]
% \begin{center}
% \includegraphics[width=10cm]{Fig2.eps}
% \end{center}
% \caption{
% High-field magnetization $M$ of LiV$_2$O$_4$ and LiV$_2$O$_{3.92}$
% powder samples at 4.2 K. $M$ of LiV$_2$O$_{3.92}$ is shifted by
% $-0.1 \mu_{\rm B}$ for clarity.}
%\end{figure}%
%%%%%%%%%%%%%%%%%%%%%%%%%%%%%%%%%%%%%%%%%%%%%%%%%%%%%%%%%%%%%
For both the systems, $M$ increases linearly up to 15 T. 
At higher fields, $M$ of LiV$_2$O$_4$ turns to increase nonlinearly,
and exhibits an inflection point at $H_{\rm m}$ = 38 T.
Such a metamagnetic behavior in LiV$_2$O$_4$ is also reported 
in the literature~\cite{Johnston}.
Magnetic moment at 45 T is about 1.1 $\rm\mu_B$ per formula unit (f.u.).
In contrast, $M$ of LiV$_2$O$_{3.92}$ increases monotonically up to 40 T with
no indication of metamagnetic behavior.
The value of $M$ at 40 T is less than 0.7 $\rm\mu_B$/f.u.,
considerably smaller than that of LiV$_2$O$_4$.

Figure 3 shows the Arrott-plot of the two systems.
%%%%%%%%%%%%%%%%%%%%%%%%%%%%%%%%%%%%%%%%%%%%%%%%%%%%%%%%%%%%
%\begin{figure}[!tbp]
% \begin{center}
% \includegraphics[width=9cm]{Fig3.eps}
% \end{center}
% \caption{
% Arrot plot for LiV$_2$O$_4$ and LiV$_2$O$_{3.92}$ at 4.2 K.}
%\end{figure}%
%%%%%%%%%%%%%%%%%%%%%%%%%%%%%%%%%%%%%%%%%%%%%%%%%%%%%%%%%%%%
The slope of $M^2$ vs. $H/M$ for LiV$_2$O$_4$ is negative, 
consistent with the metamagnetic behavior. 
On the other hand, the slope for LiV$_2$O$_{3.92}$ is positive, 
in a clear contrast to that for LiV$_2$O$_4$. 
One can therefore expect that $M$ of LiV$_2$O$_{3.92}$ increases monotonically
until its saturation value, without metamagnetic behavior even for
much higher fields.

\section{Discussion}
In this section, we argue on the possible mechanism for the metamagnetic
behavior in LiV$_2$O$_4$.
Metamagnetic behavior is usually explained as a spin-flopping
process in antiferromagnetic (AFM) ordered systems. 
In contrast, LiV$_2$O$_4$ does not order magnetically
down to the lowest temperature,
as is demonstrated by many works including 
specific heat~\cite{Kondo,Johnston99,Kaps},
$^7$Li-NMR~\cite{Kondo,Onoda,Fujiwara,Mahajan,Kaps},
muon spin relaxation~\cite{Kondo,Merrin} and
neutron diffraction~\cite{Chmaissem}.
Hence, the metamagnetic behavior observed in LiV$_2$O$_4$ is not
attributed to a spin-flopping.
Alternatively, metamagnetic behavior out of a paramagnetic state has been
observed in such cases as:
(i) Kondo-lattice systems like CeRu$_2$Si$_2$~\cite{Flouquet}, 
CeCu$_6$~\cite{Schroder,Lohneysen} or YbCuAl~\cite{Hewson},
and (ii) itinerant-electron systems close to ferromagnetic (FM) order~\cite{Goto01}
like YCo$_2$~\cite{Sakakibara} 
or TiBe$_2$~\cite{Monod}.
Notably, both the two models, Kondo-lattice and nearly-FM metal,
have been used to explain
the anomalous physical properties of LiV$_2$O$_4$.

For Kondo-lattice systems,
the precise mechanism of the metamagnetic behavior has not been
clearly understood yet. 
Nevertheless, it is widely accepted that 
the metamagnetic behavior accompanies
a qualitative change in the $f$-electron state.
For CeRu$_2$Si$_2$, where a distinct metamagnetic behavior has been
observed at $H_m$ = 7 T~\cite{Flouquet}, 
a drastic change in the magnetic response has been demonstrated by the
neutron scattering studies~\cite{Mignod};
from strong AFM fluctuations at $H < H_m$
to single-site ($q$-independent) fluctuations at $H > H_m$.
Similar change in the magnetic response has also been observed in CeCu$_6$
~\cite{Mignod}.
Furthermore, the de Haas-van Alphen effect measurements on CeRu$_2$Si$_2$
have revealed the change in its Fermi surface from
itinerant at low fields to almost localized $f$-electrons at high fields
~\cite{Takashita}.
Hence, the high-field states in these compounds are well described by 
localized $f$-electrons with single-site Kondo fluctuations.
It is also notable that 
the magnetization of Ce$_{1-x}$(La,Y)$_x$Ru$_2$Si$_2$ at high fields
is almost independent of $x$, though the metamagnetic behavior becomes broadened
with increasing $x$~\cite{Matsuhira}.
Similar behavior is also observed in Ce$_{1-x}$La$_x$Fe$_2$Ge$_2$~\cite{Sugawara} 
and Yb$_{1-x}$Y$_x$CuAl~\cite{Mattens}.
This substitution dependence is well explained 
by the single-site character of $f$-electrons at high fields,
since the single-site interactions are insensitive to disorder.
Thus, we can conclude that such a localized-electron state at high fields 
is a common feature in Kondo-lattice systems.

Here, note the qualitative difference in the magnetization of
LiV$_2$O$_4$ and  LiV$_2$O$_{3.92}$: $M$ of LiV$_2$O$_{3.92}$ at 40 T
is considerably smaller than that of LiV$_2$O$_4$.
This distinct difference in $M$ does not agree with
the high-field state expected for Kondo-lattice systems.
Therefore, the Kondo-lattice model is unlikely for LiV$_2$O$_4$ 
as far as the mechanism of the metamagnetic behavior
is concerned.
Instead, this result is likely to suggest that the
intersite correlation is still important even at high fields,
because intersite correlation should be sensitive to disorder or defects.
This indicates that the metamagnetic behavior is relevant to
the evolution of intersite FM correlation.
Thus, we turn to the next picture; itinerant-electron systems close
to FM order.

Within the itinerant-electron description, 
the Curie-Weiss susceptibility in LiV$_2$O$_4$
is attributed to the temperature dependent FM ($q \sim 0$) spin fluctuations,
not to localized moments~\cite{Moriya}.
Here, Curie constants do not necessarily agree with the effective magnetic
moment expected for a free ion, and 
a negative Weiss temperature no longer indicates
antiferromagnetic (AFM) interactions.
Several nearly-FM metals exhibit 
metamagnetic behavior as well as a maximum in $\chi$~\cite{Goto01}.
Such examples are YCo$_2$~\cite{Sakakibara}, TiBe$_2$~\cite{Monod},
and Co(S,Se)$_2$~\cite{Goto97}.
Recently, metamagnetic behavior has also been reported
in the metallic oxide Sr$_3$Ru$_2$O$_7$~ around $H$ = 5 T~\cite{Perry}. 
It is notable that this compound shows a maximum in $\chi$
and a large $\gamma \sim$ 100 mJ/Ru-mol$\cdot$K$^2$,
and is considered to be close to a FM ordering~\cite{Ikeda}.
These features of nearly-FM metals are quite similar
to those observed in LiV$_2$O$_4$ including the metamagnetic behavior.

On the other hand, it has been shown 
that LiV$_2$O$_4$ also exhibits quite contrasting behavior
to the feature for nearly-FM metals.
For example, neutron scattering experiments by Lee {\it el al.} have 
revealed the presence of AFM fluctuations at $Q \sim$ 0.6 \AA$^{-1}$
~\cite{Lee}.
In addition, spin-glass ordering has been observed in Mg- or Zn-doped
samples~\cite{Onoda,Ueda,Urano}.
These observations suggest the presence of strong AFM interaction and the
effect of the geometrical frustration, the latter of which prevents this system
from a long-range AFM ordering.
In this respect, LiV$_2$O$_4$ has been compared with 
(Y,Sc)Mn$_2$ and $\beta$-Mn~\cite{Lacroix},
where the AFM order is suppressed by the geometrical frustration,
and the resulting fluctuations enhance 
the electronic specific heat coefficients~\cite{Siga,Nakamura,Ballou}.

It should, however, be stressed that 
the temperature dependence of $\chi$ of (Y,Sc)Mn$_2$ and $\beta$-Mn
differs qualitatively from that of LiV$_2$O$_4$.
$\chi$ of the formers are almost $T$-independent 
and do not show Curie-Weiss behavior~\cite{Siga,Nakamura},
consistent with the dominant AFM ($q \gg$ 0) fluctuations.
In contrast, $\chi$ as well as the $^7$Li-Knight shift $K$ of 
LiV$_2$O$_4$ show Curie Weiss behavior, 
suggesting the importance of the FM ($q \sim$ 0) fluctuation.
Furthermore, the metamagnetic behavior observed in the present study
evidences the importance of the FM interaction in this system.
From these facts, we conclude that
both the FM ($q \sim$ 0) and the AFM ($Q \sim$ 0.6 \AA$^{-1}$)
interactions are important in LiV$_2$O$_4$.
This is similar to the case in V$_5$Se$_8$, 
where both FM ($q \sim$ 0) and 
AFM ($q \sim Q$) fluctuations are considered to
be enhanced~\cite{Kitaoka}.
This model has successfully explained the itinerant AFM behavior
and the Curie-Weiss susceptibility of V$_5$Se$_8$~\cite{Kitaoka}.

The presence of two kind of spin-fluctuations in LiV$_2$O$_4$ has been suggested
by the earlier neutron scattering study~\cite{Krimmel}
and the SCR analysis for the specific heat data~\cite{Kaps}.
Recently, Eyert {\it et al.} have performed first-principles calculations,
and have shown that AFM state is stable for LiV$_2$O$_4$
and the energy of FM state is slightly higher~\cite{Eyert}.
This result is indicative that applying high fields would reduce the
energy for the FM state, which results in a metamagnetic transition.

It is notable that the temperature where $\chi$ has a maximum is
$T_{\rm max} \simeq$ 10 K for TiBe$_2$~\cite{Polatsek}
and $T_{\rm max}$ = 16 K for Sr$_3$Ru$_2$O$_7$~\cite{Ikeda},
comparable to $T_{\rm max}$ = 16 K for LiV$_2$O$_4$,
whereas the field where metamagnetic behavior occurs is 
$H_{\rm m}$ = 6 T for TiBe$_2$~\cite{Monod} 
and $H_{\rm m}$ = 5 T for Sr$_3$Ru$_2$O$_7$~\cite{Perry},
remarkably smaller than $H_{\rm m} \simeq$ 38 T for LiV$_2$O$_4$.
This may be a result of competition of the FM and the AFM
fluctuations in LiV$_2$O$_4$ below $H_{\rm m}$.
We expect that the AFM fluctuations are 
suppressed at around $H_{\rm m}$ and the FM correlation becomes dominant
above $H_{\rm m}$.
On the other hand, FM correlation does not appear to develop
in LiV$_2$O$_{3.92}$. 
This suggests that the AFM interaction is dominant
even at high fields in this system.

Instead of the itinerant-electron picture mentioned above, 
the possibility of LiV$_2$O$_4$ as a localized-moment system
is not still ruled out.
The value of the spin-lattice relaxation rate 1/$T_1$
of the $^{51}$V ~\cite{Onoda}
and of the $^7$Li nuclei ~\cite{Mahajan} at high temperatures
are consistent with those for a localized-moment system.
The evolution of localized electrons has been suggested theoretically
by the splitting of the $t_{2g}$ orbital into almost localized $A_{1g}$ 
and conductive $E_{g}$ orbitals due to a trigonally distortion in the VO$_6$
octahedra and the Coulomb interactions~\cite{Anisimov}.
In this case, the most probable explanation for the metamagnetic behavior
is the evolution of FM interaction due to 
the double exchange mechanism.
Here, the FM interaction evolves via the Hund coupling between
local moments and conduction electrons,
while AFM interaction can also arise via the exchange interaction
between local moments~\cite{Onoda,Lacroix,Anisimov}.
This would also result in the situation where both the FM and the AFM interactions
are important.

Although our results do not clarify whether the $d$-electron state
in LiV$_2$O$_4$ is itinerant or localized, 
the absence of metamagnetic behavior in LiV$_2$O$_{3.92}$
seems to be of particular importance.
The reduced $M$ in this system implies 
that the FM interaction is rapidly reduced by a slight disorder or defects.
This may be a hint for the origin of the FM interaction in LiV$_2$O$_4$.
Since the $d$-electrons in LiV$_2$O$_4$ involve 
degrees of freedom of orbital and charge quanta,
orbital and/or charge order would also be suppressed
by the geometrical frustration.
This multiply frustrated state can be easily lifted by a small defect.
The FM correlation in LiV$_2$O$_4$ may be a result of the competition  
of the multiply frustrated state.
To elucidate the electronic state and 
the mechanism for the evolution of the FM
correlation in LiV$_2$O$_4$,
more systematic experiments as well as theoretical studies should be
performed.  

\section{Conclusion}
We have reported a metamagnetic behavior in LiV$_2$O$_4$ around $H_m$ = 38 T and
absence of metamagnetism in LiV$_2$O$_{3.92}$ up to 40 T.
Arrot plot of the former shows the characteristic curve for a metamagnetic behavior,
while that for the latter has a positive curve.
This implies that metamagnetic behavior would not appear for LiV$_2$O$_{3.92}$
even at much higher fields.
Moreover, high-field magnetic moment of this system is significantly smaller
from that of LiV$_2$O$_4$.
This qualitative difference in the magnetization of the two samples 
is of particular importance to consider the mechanism of the metamagnetic behavior.

Metamagnetic behavior out of a paramagnetic state has been observed in two cases:
Kondo-lattice systems and itinerant-electron systems with strong ferromagnetic
spin-fluctuations.
For the case of Kondo-lattice systems, the high-field state is described by
localized electrons with single-site Kondo fluctuations.
Consequently, the magnetic moments at high fields are insensitive to a slight 
disorder or defects for Kondo-lattice systems.
For the present case, 
the significant difference in the magnetization of LiV$_2$O$_4$
and LiV$_2$O$_{3.92}$ does not agree with the local moment state.
The Kondo-lattice model is therefore unlikely as far as the
mechanism for the metamagnetic behavior is concerned.
Instead, these results favor the evolution of intersite ferromagnetic
spin-correlation at high fields for the origin of metamagnetic behavior 
in LiV$_2$O$_4$.

This may lead to the interpretation that LiV$_2$O$_4$ is a nearly-ferromagnetic
metals, as is suggested by Fujiwara {\it et al.} from their $^7$Li-NMR
experiments~\cite{Fujiwara}.
In fact, many itinerant-electron systems with strong ferromagnetic spin-fluctuations
exhibit metamagnetic behavior by fields~\cite{Goto01}.
For LiV$_2$O$_4$, however, neutron scattering experiments have revealed
the presence of antiferromagnetic fluctuations at low temperature~\cite{Lee}.
From these facts, we conclude that both the ferromagnetic and the antiferromagnetic
interactions are important in LiV$_2$O$_4$, and the former is strongly enhanced
by external fields.

Microscopic origin of the ferromagnetic correlation is unclear. 
The qualitative difference in the high-field magnetizations of LiV$_2$O$_4$
and LiV$_2$O$_{3.92}$ is likely to suggest that the effect of
geometrical frustration for spin, orbital, and charge order 
is relevant to the ferromagnetic interaction.
The other possibility for the ferromagnetic correlation would be
the double exchange mechanism.

\begin{figure}[!bp]
 \begin{center}
 \includegraphics[width=10cm]{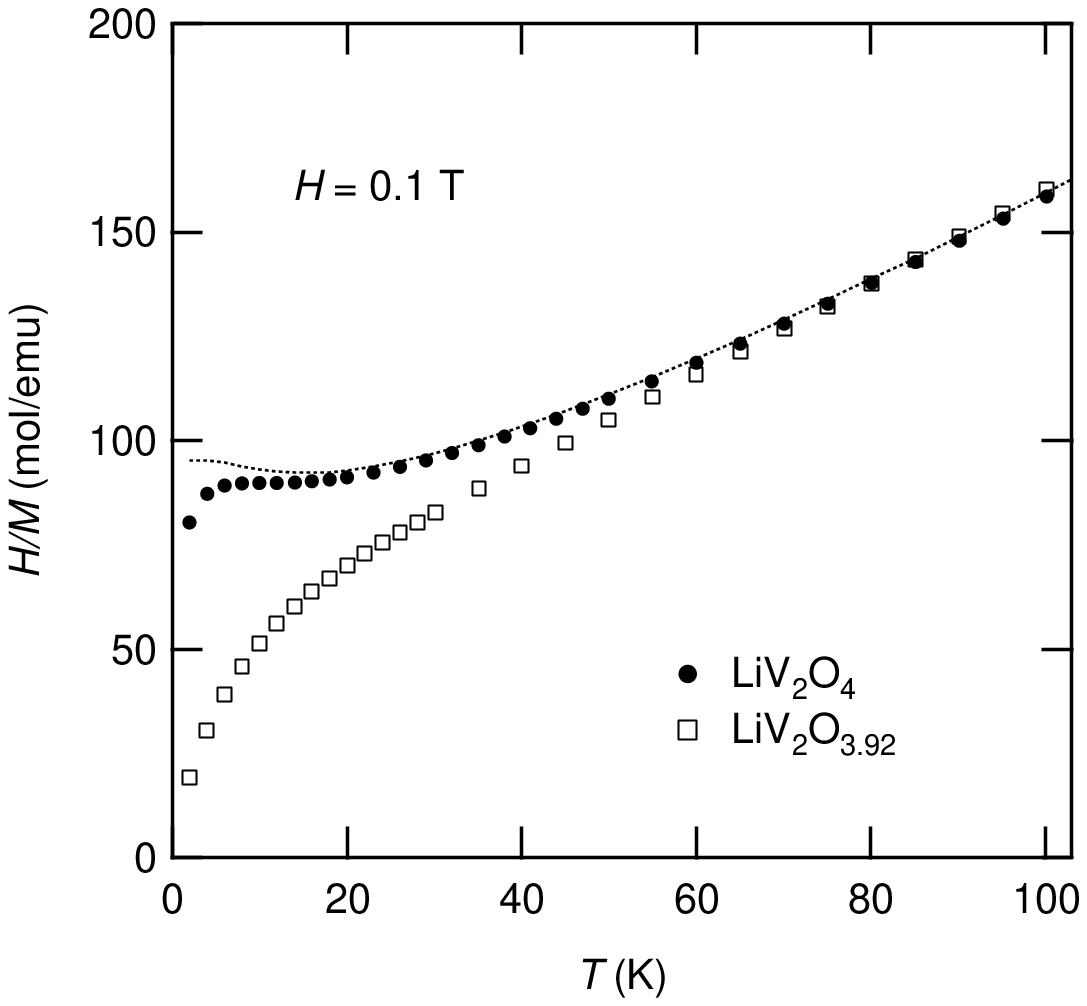}
 \end{center}
 \caption{
 Inverse of the magnetic susceptibility $1/\chi$ of LiV$_2$O$_4$ and LiV$_2$O$_{3.92}$
 powder samples. Dotted line represents the corrected data of LiV$_2$O$_4$ 
 with the Curie term
 subtracted.}
\end{figure}%
\begin{figure}[!tbp]
 \begin{center}
 \includegraphics[width=10cm]{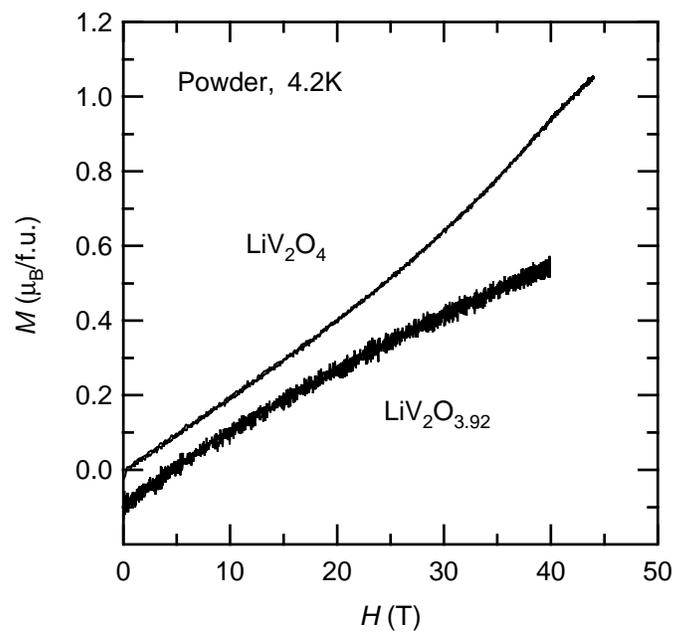}
 \end{center}
 \caption{
 High-field magnetization $M$ of LiV$_2$O$_4$ and LiV$_2$O$_{3.92}$
 powder samples at 4.2 K. $M$ of LiV$_2$O$_{3.92}$ is shifted by
 $-0.1 \mu_{\rm B}$ for clarity.}
\end{figure}%
\begin{figure}[!tbp]
 \begin{center}
 \includegraphics[width=10cm]{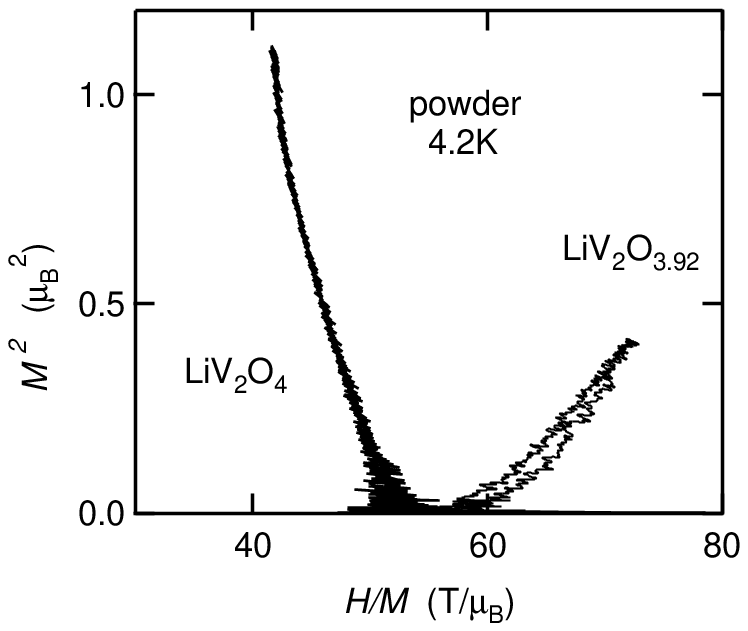}
 \end{center}
 \caption{
 Arrot plot for LiV$_2$O$_4$ and LiV$_2$O$_{3.92}$ at 4.2 K.}
\end{figure}%

\section*{References}
\begin{thebibliography}{99}
\bibitem{Kondo} Kondo S, Johnston D C, Swenson C A, Borsa F, Mahajan A V, Miller L L,
Gu T, Goldman A I, Maple M B, Gajewski D A, Freeman E J, Dilley N R, Dickey R P, Merrin J,
Kojima K, Luke G M, Uemura Y J, Chmaissem O and Jorgensen J D 1997 {\it Phys. Rev. Lett.}
{\bf 78} 3729
\bibitem{Onoda} Onoda M, Imai H, Amako Y and Nagasawa H 1997 {\it Phys. Rev.} B
{\bf 56} 3760
\bibitem{Ueda} Ueda Y, Fujiwara N and Yasuoka H 1997 {\it J. Phys. Soc. Jpn.}
{\bf 66} 778
\bibitem{Urano} Urano C, Nohara M, Kondo S, Sakai F, Takagi H, Shiraki T and 
Okubo T 2000 {\it Phys. Rev. Lett.} {\bf 85} 1052
\bibitem{Fujiwara} Fujiwara N, Ueda Y and Yasuoka H 1997 {\it Physica} B 
{\bf 237-238} 59\\
Fujiwara N, Yasuoka H and Ueda Y 1998 {\it Phys. Rev.} B {\bf 57} 3539\\
Fujiwara N, Yasuoka H and Ueda Y 1999 {\it Phys. Rev.} B {\bf 59} 6294
\bibitem{Mahajan} Mahajan A V, Sala R, Lee E, Borsa F, Kondo S and Johnston D C
1998 {\it Phys. Rev.} B {\bf 57} 8890
\bibitem{Kawakami} Kawakami K, Sakai Y and Tsuda N 1986 {\it J. Phys. Soc. Jpn.}
{\bf 55} 3174
\bibitem{Muhtar} Muhtar, Takagi F, Kawakami K and Tsuda N 1988
{\it J. Phys. Soc. Jpn.} {\bf 57} 3119
%\bibitem{Moriya} Moriya T 1985 {\it Spin Fluctuations in Itinerant Electron
%Magnetism} Springer
\bibitem{Krimmel} Krimmel A, Loidl A, Klemm M, Horn S and Schober H 1999
{\it Phys. Rev. Lett.} {\bf 82} 2919\\
Murani A P 2000 {\it Phys. Rev. Lett.} {\bf 85} 3981\\
Krimmel {\it et al.} 2000 {\it Phys. Rev. Lett.} {\bf 85} 3982
\bibitem{Lacroix} Lacroix C 2001 {\it Can. J. Phys.} {\bf 79} 1469
\bibitem{Nakajima} Nakajima Y, Amamiya Y, Ohnishi K, Terasaki I,
Maeda A and Uchinokura K 1991 {\it Physica} C {\bf 185-189} 719
\bibitem{Johnston} Johnston D C 2000 {\it Physica} B {\bf 281-282} 21
\bibitem{Johnston99} Johnston D C, Swenson C A and Kondo S 1999
{\it Phys. Rev.} B {\bf 59} 2627
\bibitem{Kaps} Kaps H, Brando M, Trinkl W, B\"{u}ttgen N, Loidl A,
Scheidt E-W, Klemm M and Horn S 2001 {\it J. Phys.: Condens. Matter}
{\bf 13} 8497
\bibitem{Merrin} Merrin J, Fudamoto Y, Kojima K M, Larkin M, Luke G M,
Nachumi B, Uemura Y J, Kondo S and Johnston D C 1998
{\it J. Mag. Mag. Mater.} {\bf 177-181} 799
\bibitem{Chmaissem} Chmaissem O, Jorgensen J D, Kondo S and Johnston D C 1997
{\it Phys. Rev. Lett.} {\bf 79} 4866
\bibitem{Flouquet} Flouquet J, Kambe S, Regnault L P, Haen P, Brison J P,
Lapierre F and Lejay P 1995 {\it Physica} B {\bf 215} 77
\bibitem{Schroder} Schr\"{o}der A, Schlager H G and L\"{o}hneysen H v
1992 {\it J. Mag. Mag. Mater.} {\bf 108} 47
\bibitem{Lohneysen} L\"{o}hneysen H v, Schlager H G and Schr\"{o}der A
1993 {\it Physica} B {\bf 186} 590
\bibitem{Hewson} Hewson A C and Rasul J W 1983 {\it J. Phys.} C {\bf 16} 6799
\bibitem{Goto01} For a review,
Goto T, Fukamichi K and Yamada H 2001 {\it Physica} B {\bf 300} 167
\bibitem{Sakakibara} Sakakibara T, Goto T, Yoshimura K and Fukamichi K 1990
{\it J. Phys.: Condens. Matt.} {\bf 2} 3381
\bibitem{Monod} Monod P, Felner I, Chouteau G and Shaltiel D 1980
{\it J. Physique} {\bf 41} L511
\bibitem{Mignod} Rossat-Mignod J, Regnault L P, Jacoud J L, Vettier C, Lejay P,
Flouquet J, Walker E, Jaccard D and Amato A 1988 {\it J. Mag. Mag. Mater.}
{\bf 76-77} 376
\bibitem{Takashita} Takashita M, Aoki H, Terashimra T, Uji S,
Maezawa K, Settai R and \={O}nuki Y 1996 {\it J. Phys. Soc. Jpn.}
{\bf 65} 515
\bibitem{Matsuhira} Matsuhira K, Sakakibara T, Amitsuka H, Tenya K, Kamishima K, 
Goto T and Kido G 1997 {\it J. Phys. Soc. Jpn.} {\bf 66} 2851
\bibitem{Sugawara} Sugawara H, Namiki T, Yuasa S, Matsuda T D, Aoki Y,
Sato H, Mushnikov N, Hane S and Goto T 2000 {\it Physica} B {\bf 281-282} 69
\bibitem{Mattens} Mattens W C M, de Ch\^{a}tel P F, Moleman A C and
de Boer F R 1979 {\it Physica} B {\bf 96} 138 
\bibitem{Moriya} Moriya T 1985 {\it Spin Fluctuations in Itinerant
Electron Magnetism}, Springer-Verlag
\bibitem{Goto97} Goto T, Shindo Y, Takahashi H and Ogawa S 1997
{\it Phys. Rev.} B {\bf 56} 14019
\bibitem{Perry} Perry R S, Galvin L M, Grigera S A, Capogna L, Schofield A J,
Mackenzie A P, Chiao M, Julian S R, Ikeda S I, Nakatsuji S, Maeno Y and 
Pfleiderer C 2001 {\it Phys. Rev. Lett.} {\bf 86} 2661
\bibitem{Ikeda} Ikeda S I, Maeno Y, Nakatsuji S, Kosaka M and Uwatoko Y 2000
{\it Phys. Rev.} B {\bf 62} R6089
\bibitem{Lee} Lee S-H, Qiu Y, Broholm C, Ueda Y and Rush J J 2001
{\it Phys. Rev. Lett.} {\bf 86} 5554
\bibitem{Siga} Shiga M, Fujisawa K and Wada H 1993
{\it J. Phys. Soc. Jpn.} {\bf 62} 1329
\bibitem{Nakamura} Nakamura H, Yoshimoto K, Shiga M, Nishi M and Kakurai K 1997
{\it J. Phys.: Condens. Matter} {\bf 9} 4701
\bibitem{Ballou} Ballou R, Leli\`{e}vre-Berna E and F\aa{}k B 1996
{\it Phys. Rev. Lett.} {\bf 76} 2125
\bibitem{Kitaoka} Kitaoka Y and Yasuoka H 1980 {\it J. Phys. Soc. Jpn.}
{\bf 48} 1460
\bibitem{Eyert} Eyert V, H\"{o}ck K-H, Horn S, Loidl A and Riseborough P S 1999
{\it Europhys. Lett.} {\bf 46} 762
\bibitem{Polatsek} Polatsek G and Zevin V 1988 {\it J. Mag. Mag. Mater.}
{\bf 73} 205
\bibitem{Anisimov} Anisimov V I, Korotin M A, Z\"{o}lfl M, Pruschke T,
Le Hur K and Rice T M 1999 {\it Phys. Rev. Lett.} {\bf 83} 364
\end{thebibliography}
\end{document}